\documentclass[12pt,epsf]{article}

\usepackage[dvips]{graphicx}

\newcommand{\beq}{\begin{equation}}
\newcommand{\eeq}{\end{equation}}
\newcommand{\bea}{\begin{eqnarray}}
\newcommand{\eea}{\end{eqnarray}}
\usepackage{bm}

\begin{document}
\thispagestyle{empty}
\vspace*{-15mm}
{\bf OCHA-PP-340}\\

\vspace{15mm}
\begin{center}
{\Large\bf
Construction of a Model of Monopolium and \\ its Search via Multi-Photon Channels at LHC
}

\vspace{7mm}

\baselineskip 18pt
{\bf Neil  D. Barrie$^{1}$, Akio Sugamoto $^{2, 3}$ and Kimiko Yamashita$^{2, 4}$}
\vspace{2mm}

{\it
$^1$ ARC Center of Excellence for Particle Physics at the Terascale, \\
School of Physics, The University of Sydney, NSW 2006, Australia} \\
{\it $^2$Department of Physics, Graduate School of Humanities and Sciences, \\
   Ochanomizu University, Tokyo 112-8610, Japan}\\
{\it $^3$Tokyo Bunkyo SC, the Open Universtiy of Japan, Tokyo 112-0012, Japan}\\
{\it $^4$Program for Leading Graduate Schools, \\
Ochanomizu University, Tokyo 112-8610, Japan} \\

\vspace{10mm}
\end{center}
\begin{center}
\begin{minipage}{14cm}
\baselineskip 16pt
\noindent
\begin{abstract}%
A model of monopolium is constructed based on an electromagnetic dual formulation of Zwanziger and lattice gauge theory.  
    
To cope with the strong coupling nature of the magnetic charge, for which the monopole is confined, $ U(1) $ lattice gauge theory is applied.
The monopole is assumed to have a finite-sized inner structure based on a 't Hooft-Polyakov like solution in which the magnetic charge is uniformly distributed on the surface of a sphere. 
The monopole and antimonopole potential becomes linear plus Coulomb outside the sphere and is constant inside.  
 
Numerical estimation gives two kinds of solutions: One which has a small binding energy, and hence the para-($ J=0 $) and ortho-($ J=1 $) monopoliums have degenerate masses. For the parameter choices considered, they both have $ \mathcal{O}(1-10) $ TeV masses and are very short-lived.  The other solution has a small monopole mass and large binding energy, with an illustrative example of parameter choices giving a 750 GeV para-monopolium and 1.4 TeV ortho-monopolium.  The production rate of the former is one order of magnitude smaller than the announced enhancement, but they may be the target of future LHC searches and the 100 TeV colliders.
\end{abstract}

\end{minipage}
\end{center}


\section{Introduction}
Last year, an interesting enhancement was reported at 750 GeV by the ATLAS \cite{ATLAS} and CMS \cite{CMS} groups at LHC.\footnote{At  ICHEP 2016 (August 3-10), this enhancement was reported to disappear after adding the new data collected by ATLAS and CMS in 2016.  Even if this paper is triggered by the enhancement, we study the monopolium in general, so that the contents are not affected by whether it may exist or not.}
The enhancement $X$ was observed in the $\gamma \gamma$ channel and was not observed in the other di-boson channels such as ${ZZ}$, ${W^+} $ and ${W^-}$: 
\begin{eqnarray}
\sigma(p+p \to X \to \gamma \gamma)=10 \pm 3~\mbox{fb}~\mbox{(ATLAS)} ~\mbox{and}~ 6 \pm 3 ~\mbox{fb}~\mbox{(CMS)}, 
\end{eqnarray}
at 13 TeV (Run2) experiment, but no such enhancement was reported at 8~TeV (Run1). Upon combining the ATLAS and CMS data and including 8~TeV data the following cross-section estimate is obtained \cite{Buttazzo:2015txu}:
\begin{equation}
\sigma (pp\to X \to \gamma\gamma) \approx  (4.6\pm1.2)~ {\rm fb}.
\label{combined}
\end{equation}
The total width $\Gamma_\mathrm{tot}$ has not been well determined, but appears to be large $ \sim45 $ GeV.  Various models have been proposed for $X$ \cite{750 GeV models}, but if $X$ couples exclusively to $\gamma \gamma$ with a large coupling constant, then it may be possible to consider the enhancement as a monopolium, or a monopole-antimonopole dumbbell as described by Nambu \cite{Nambu}.  A recent paper by M. Yamada {\it  et al.}  \cite{Yamada et al.} considers a similar idea, but they identify the enhancement as a magnetic Higgs particle where it is dual to the electric Higgs particle in a hidden $\mbox{U(1)}_H$ gauge theory.  Ours is more directly a bound state system of a monopole and an antimonopole.  

The detection of monopolium has already been studied by L. I. Epele {\it et al.}, aiming for observation at Tevatron and LHC \cite{Epele1, Epele2}. The main difference between their paper and the following analysis is that we adopt a strong coupling expansion in lattice gauge theory and accordingly a linear term is included in the monopole-antimonopole potential. Therefore, the large binding energy case can be studied. 
  
This work is motivated by papers ~\cite{Neil1, Neil2} written by one of the authors (N. D. B.) on explaining the diphoton excess via photon fusion production of $ X $, in particular where $ X $ is identified as a leptonium of highly charged leptons having a charge $Q=(5-7)~e$.  Higher charge can be naturally understood in the context of monopoles due to the large magnetic charge, which is opposite to the small electric charge. Also, the stability of the monopole is less crucial than the lepton, since our monopole is confined and can't exist individually.
  
The difficulty of highly charged particles is the estimation of the potential between them.  One photon exchange is not enough, and so we will use lattice gauge theory with a finite lattice constant $a$, in which a strong coupling expansion is possible \cite{lattice gauge theory1, lattice gauge theory2, lattice gauge theory3}.  We will apply this to the magnetic $ U(1) $ part of the manifestly electromagnetic dual formulation of Zwanziger \cite{Zwanziger1, Zwanziger2}.  In lattice gauge theory, however, there exists another difficulty, that is, $ U(1) $ gauge theories are not well defined; a first-order phase transition exists between the weak coupling perturbative region and the strong coupling confinement region, and so the continuum limit of $a \to 0$ can't be taken properly \cite{Creutz-Jacobs-Rebbi}.
On the other hand, the continuum limit is properly taken for $ SU(2) $ \cite{Creutz} and other asymptotically free gauge theories.  So, we consider that the welcoming non-Abelian structure reveals when we go inside the finite sized $ U(1) $ monopole, that is, at some point of taking $a \to 0$, the gauge group is expected to be enhanced from $ U(1) $ to $ SU(2) $ and the 't Hooft-Polyakov~\cite{monopole1, monopole2, monopole3} like inside structure will appear.  Even in the 't Hooft-Polyakov monopole,  the $ U(1) $ magnetic charge is unfortunately located at the origin as a point-like singularity.  To relax this situation, we look for a solution in which the $ U(1) $ magnetic charge is distributed nonlocally.  Fortunately, there exists a solution in which the magnetic charge is distributed uniformly on the surface of a sphere with radius $R$.  Inside the sphere ($r<R$) there is no magnetic force, and so the potential becomes flat, while the potential between the monopole and antimonopole becomes linear plus Coulomb outside the sphere. 

We estimate the monopolium masses and wave functions using this potential.  In this model, the monopole and antimonopole are confined like quark and antiquark, so that they are not observed individually.
  
If the monopolium is found and the world of monopoles is opened in reality, then various interesting phenomena are expected. One of these is that the neutrino mass as well as possible lepton and baryon number violation effects may be attributed to the monopolium or the monopole-antimonopole dumb bell system \cite{sugamoto}. 

\section{Zwanziger's manifestly dual formulation of gauge theory} 
The manifestly dual formulation of $ U(1) $ gauge theory by Zwanziger \cite{Zwanziger1, Zwanziger2} is  given by the following action,   
\begin{eqnarray}
S^{\rm{ZW}}&=&\int d^4 x \left[ \left(\frac{1}{2} \eta^{\mu} \eta^{\lambda}
               \sqrt{-g} g^{\nu\rho}\right) \nonumber \right. \\
&\times& \left(F_{\mu\nu} F_{\lambda\rho} + G_{\mu\nu} G_{\lambda\rho}+ F_{\mu\nu} \tilde{G}_{\lambda\rho} - G_{\mu\nu}\tilde{F}_{\lambda\rho} \right)  \nonumber \\
&+& \left. \sum_i \overline{\psi_i} \gamma^{\mu}( iD_{\mu}-m_i) \psi_i \right]
   \end{eqnarray}
Here, a constant unit vector $\eta^{\mu}$, denoting the direction of Dirac strings \cite{Dirac1, Dirac2, Dirac3} is displayed in parallel along the space-like direction, and
 \begin{eqnarray}
              iD_{\mu}&=&i \partial_{\mu}- e_i A_{\mu} -g_i B_{\mu}, \\
            F_{\mu\nu}&=&\partial_{\mu}A_{\nu}-\partial_{\nu}A_{\mu}, ~~
           G_{\mu\nu}=\partial_{\mu}B_{\nu}-\partial_{\nu}B_{\mu}, \\
           \tilde{F}_{\mu\nu}&=&\frac{1}{2} \varepsilon_{\mu\nu\lambda\rho}
                               F^{\lambda\rho}, ~~\tilde{G}_{\mu\nu}=\frac{1}{2} \varepsilon_{\mu\nu\lambda\rho}
                               G^{\lambda\rho}, 
\end{eqnarray}
where fermions with electric charge $e_i$ and magnetic charge $g_i$ are introduced, and $\varepsilon^{0123}=1$.  We can add scalars if necessary.  From the consistency condition of the finite Lorentz transformation, the following Dirac (or correctly Schwinger type) quantization condition \cite{Dirac1, Dirac2, Dirac3} appears
\begin{eqnarray}
e_ig_j-g_ie_j= 4\pi N_{ij},
\end{eqnarray}
where $N_{ij}$ is an integer.
Here, we consider a flat space-time $g_{\mu\nu}=\eta_{\mu\nu}=(1, -1, -1, -1)$ and $\varepsilon^{0123}=-\varepsilon_{0123}=1$.  If we want to obtain the dual gravity theory, then consider that the metric $g_{\mu\nu}(x)$ is formed as a collective excitation of a pair of gauge bosons, $(B_{\mu}(x)B_{\nu}(x))$ or $(A_{\mu}(x)A_{\nu}(x))$ \cite{Noguchi-Sugamoto}.  In the Zwanziger formulation, the degrees of freedom are doubled by the introduction of electric and magnetic vector potentials, but are halved by the projection to the $\eta^{\mu}$ direction.  Such a formulation in a non-Abelian gauge theory is more complicated but is possible~\cite{Zwanziger3, Zwanziger4, Zwanziger5}.

If the axial gauge is taken,
\begin{eqnarray}
\eta^{\mu}A_{\mu}(x)=\eta^{\mu}B_{\mu}(x)=0,
\end{eqnarray}
no ghost fields appear, and the Feynman rules are obtained as follows~:
\begin{eqnarray}
\langle A_{\mu} A_{\nu} \rangle (k)&=& \frac{-i}{k^2 + i \varepsilon} \left( g_{\mu\nu} - \frac{k_{\mu}\eta_{\nu}+k_{\nu}\eta_{\mu}}{k \cdot \eta}+\eta^2 \frac{k_{\mu}k_{\nu}}{(k \cdot \eta)^2} \right), \\
\langle B_{\mu} B_{\nu} \rangle (k)&=& \frac{-i}{k^2 + i \varepsilon} \left( g_{\mu\nu} - \frac{k_{\mu}\eta_{\nu}+k_{\nu}\eta_{\mu}}{k \cdot \eta} +\eta^2 \frac{k_{\mu}k_{\nu}}{(k \cdot \eta)^2}\right), \\
\langle A_{\mu} B_{\nu} \rangle (k) &=&-\langle B_{\mu} A_{\nu} \rangle  (k)= \frac{-i}{k^2 + i \varepsilon} \varepsilon_{\mu\nu\rho\sigma} \frac{\eta^{\rho}k^{\sigma}}{k \cdot \eta}.
\end{eqnarray}
The kinetic terms of the gauge field are complicated and depend on $\eta^{\mu}$, but  the $2 \times 2$ matrix form of the propagators is simple and satisfies
\begin{eqnarray}
\hat{D}^{ab\mu\nu} \langle V_{b\nu} V_{c\lambda} \rangle = \delta^a_c \eta^{\mu}_{\lambda},
\end{eqnarray} 
where $V^1_{\mu}=A_{\mu}$, $V^2_{\mu}=B_{\mu}$, and $\hat{D}^{ab\mu\nu}$ is the differential operator for the gauge fields in the action.\footnote{Propagators that Zwanziger derived have no $\eta^2 \frac{k_{\mu}k_{\nu}}{(k \cdot \eta)^2}$ term in the numerator \cite{Zwanziger2}. These propagators are for a gauge-fixing Lagrangian ${\cal L}_G=-\frac{1}{2} \left[ (\partial_{\mu}(\eta \cdot  A) )^2+ (\partial_{\mu}(\eta \cdot B) )^2 \right]$.  Two kinds of propagators with or without $\eta^2 \frac{k_{\mu}k_{\nu}}{(k \cdot \eta)^2}$ term are, of course, equivalent for the on-shell amplitudes, if the Ward identity is taken into account.}

\section{Production and decay of monopolium at LHC}
We assume that there is only one kind of spin 1/2 monopole $M$ in our world, that is electrically neutral, having magnetic charge $g$.  We have, of course, a number of magnetically neutral spin 1/2 fields having electric charge $Q=\pm e$  ($e$ is the unit of the electron charge) and the monopole satisfies the Schwinger quantization condition with these charged particles.  This gives
\begin{eqnarray}
g=\frac{4\pi}{e} N ~~(N=\mbox{integer}).  
\end{eqnarray}
We denote the mass of the monopole simply as $m$, and assume that the monopole $M$ and its antimonopole $\overline{M}$ form
a $J=0$ bound state $X_0=(M\overline{M})$ ($^1S_0$, para-monopolium) and a $J=1$ bound state of $X_1=(M\overline{M})$ ($^3S_1$, ortho-monopolium) via the exchange of magnetic photons, which we will denote as $\gamma_M$ and are described by the $B_{\mu}$ fields. The mass of the monopolium so obtained will be denoted generally as $m_X$, or specifically $m_{X_0}$ and $m_{X_1}$.

After obtaining the energy and wave function of the monopolium, we can estimate the various partial decay widths $\Gamma(X_0 \to f)$ for $X_0$ or $\Gamma(X_1 \to f)$ for $X_1$, and the total decay widths $\Gamma_\mathrm{tot} (X_0 \to all~f's)$ or $\Gamma_\mathrm{tot} (X_1 \to all~f's)$, where $f$ denotes a final state.

In this section we use perturbation theory even for the strong coupling monopole.  Our strategy is to store the  effects into the bound state problem or the estimation of the monopole and antimonopole potential, and study the multi-photon emission processes perturbatively.  This is because, if we fix the number of emitted photons to $n$, then even if photons convert to magnetic photons and couple to the monopolium strongly, the order of coupling is $O(g^n)$, so that perturbative calculation at this order can be applied. This is valid, however, under the restriction that radiative corrections due to monopole loops are ignored. 

\subsection{Production cross section}
At LHC, to produce $X$ we consider that a parton $a$ inside one proton collides with parton $b$ inside another proton to produce $X$ as a resonance scattering, where the parton can be anything produced from the proton, namely, quark, gluon, or photon.  To do this we introduce the parton distribution function inside the proton $f_{a/p}(x)$, where $x$ is the momentum fraction. Then the production cross section reads
\begin{eqnarray}
&&\sigma(pp \to a+b \to X \to f)  \nonumber \\
&&=\int_0^1 dx_1 \int_0^1 dx_2 f_{a/p}(x_1) f_{b/p}(x_2) \hat{\sigma}(\hat{s}) (a+b \to X \to f) \times (2, \mbox{if}~ a \ne b), \\
&&=\int_0^1 d\tau \int_{\tau}^1 \frac{dx_1}{x_1} f_{a/p}(x_1) f_{b/p}(\tau/x_1) \hat{\sigma}(\hat{s}) (a+b \to X \to f) \times (2, \mbox{if}~ a \ne b).
\end{eqnarray}
 Here, the patrons are labeled $a$ and $b$, and $\hat{\sigma}$ is the cross section of the two colliding patrons, $a$ and $b$, with the center of mass (CM) energy $\sqrt{\hat{s}}$, $\tau\equiv x_1x_2$, and $\hat{s}=\tau s$, where $\sqrt{s}$ is the CM energy of the colliding protons. 

If we introduce the ``parton luminosity" for $a$ and $b$ as $d{\cal L}_{ab}(\tau)/d\tau$, and define it by \cite{Barger-Phillips}
\begin{eqnarray}
\frac{d {\cal L}_{ab}(\tau)}{d\tau} \equiv \frac{1}{c_ac_b}\int_{\tau}^1 \frac{dx_1}{x_1} ~f_{a/p}(x_1) f_{b/p}(\tau/x_1)\times (2, ~\mbox{if}~ a \ne b),
\end{eqnarray}
where $c_a$ and $c_b$ are the color degrees of freedom for $a$ and $b$~ $(c_a= 3$ and $8$ for quark and gluon, respectively), then, the cross section becomes
\begin{eqnarray}
&&\sigma(pp \to a+b \to X \to f) \nonumber \\
&&=\int_0^1 d\tau ~\frac{d {\cal L}_{ab}(\tau)}{d\tau} ~ c_ac_b ~\hat{\sigma}(\hat{s}) (a+b \to X \to f).
\end{eqnarray}
If we replace $\hat{\sigma}$ by the resonance scattering formula, or its small width approximation \footnote{The spin degree of freedom $2s+1$ is $2s=2$ for photon and gluon.} ,
\begin{eqnarray}
\hat{\sigma}(\hat{s}) (a+b \to X \to f)= 32 \pi  \frac{(2J_X+1)F}{c_ac_b(2s_a+1)(2s_b+1)}\frac{\Gamma(X \to a+b) \Gamma(X \to f)}{(\hat{s}-m^2_X)^2 + m_X^2 \Gamma_\mathrm{tot}^2} \\
\approx 32 \pi  \frac{(2J_X+1)F}{c_ac_b(2s_a+1)(2s_b+1)}\frac{\Gamma(X \to a+b) \Gamma(X \to f)}{m_X\Gamma_\mathrm{tot}} \pi \delta(\hat{s}-m_X^2).
\end{eqnarray}
From these, we have
\begin{eqnarray}
&&\sigma(pp \to a+b \to X \to f)  \nonumber \\
&&=\left[ \frac{1}{s} \frac{d {\cal L}_{ab}(\tau)}{d\tau}\right]_{\tau=\frac{m_X^2}{s}} \times \left[  \frac{32 \pi^2 (2J_X+1)F}{(2s_a+1)(2s_b+1)} \frac{\Gamma(X \to a+b) \Gamma(X \to f)}{m_X\Gamma_\mathrm{tot}}\right], 
\end{eqnarray}
where the factor $F= 1$ for $a=b$, but is $\frac{1}{2}$ for $a \ne b$, since $a+b$ is originally the initial state, but it is converted to the final state in the formula. 

Now the production cross section is expressed as the product of two factors, the first is given by the parton luminosity function and $s^{-1}$, while the second one is a model-dependent number determined by the decay widths and mass of the resonance $X$.  If we understand that $s^{-1}=(13~\mbox{TeV})^{-2}$ and $(8 ~\mbox{TeV})^{-2}$ correspond to $2.4 ~\mbox{pb}$ and $6.2 ~\mbox{pb}$, respectively, in order to obtain 3 - 6 fb (to be consistent with Eq. (\ref{combined})), we have to find the suppression factor of $10^{-3}$ from the product of $\frac{d {\cal L}_{ab}(\tau)}{d\tau},  \Gamma(X \to a+b)/m_X$, and $\mathrm{Br}(X \to f)$.  This is one of the reasons why it is difficult to understand the 750 GeV enhancement.
\subsection{Decay rate}
The decay rate of $X_0 \to 2 \gamma$ can be estimated from $\sigma_0(M+\overline{M} \to 2\gamma)$ and the value of the wave function of relative motion of the monopole and antimonopole at the distance $ R $, $\psi_{^1S_0}(r=R)$, that is
\begin{eqnarray}
\Gamma(X_0 \to \gamma\gamma)=4 \vert \psi_{^1S_0}(R) \vert^2 \sigma_0(M+\overline{M} \to \gamma + \gamma) v_{rel},
\end{eqnarray}
since our model of the monopole has a finite size $R$, $\vert \psi_{^1S_0}(R) \vert^2$ gives the probability of finding the monopole and antimonopole at a relative distance $ R $ in the bound state, where the collision occurs.  The factor 4 $\sigma_0 v_\mathrm{rel}$ gives the reaction rate (the factor 4 comes from the possible combination of monopole and antimonopole spin states). 

Similarly, the decay rate of $X_1 \to 3 \gamma$ is estimated from $\sigma_1(M+\overline{M} \to 3\gamma)$ and the value at $r=R$ of the wave function of the ortho-monopolium $\psi_{^3S_1}(r=R)$, that is 
\begin{eqnarray}
\Gamma(X_1 \to 3\gamma)=\frac{4}{3} \vert \psi_{^3S_1}(R) \vert^2 \sigma_1(M+\overline{M} \to 3\gamma) v_\mathrm{rel}.
\end{eqnarray}

To estimate $\sigma_0(M+\overline{M} \to 2\gamma)$ or $\sigma_1(M+\overline{M} \to 3\gamma)$, we first consider the process of emission of two or three magnetic photons, and convert the emitted magnetic photons to the electric photons.  The estimation is done by using the perturbation theory, since the final number of emitted photons is restricted to two or three.  

In the calculation, the polarization vector of the magnetic photon $\epsilon_M(k)^{\mu}$, has to be converted to the usual electric photon's polarization $\epsilon_{\gamma}(k)^{\nu}$, by using the off-diagonal propagator $\langle B_{\mu} A_{\mu} \rangle (k)$, that is 
\begin{eqnarray}
\epsilon_M(k)^{\mu}=\varepsilon^{\mu\nu\rho\sigma} \epsilon_{\gamma}(k)_{\nu} \frac{\eta_{\rho} k_{\sigma}}{(k \cdot \eta)}. \label{replacement of polarizations}
\end{eqnarray}
The polarization sum of photons, $\sum_{pol}\epsilon_{\gamma}(k)^{*\lambda}\epsilon_{\gamma}(k)^{\rho}=-g^{\lambda\rho}$, is performed, then we have
\begin{eqnarray}
\sum_{\gamma's~\mathrm{pol.}}\epsilon_M(k)^{*\mu}\epsilon_M(k)^{\nu}
=g^{\mu\nu}\left(-1+\frac{\eta^2k^2}{(k \cdot \eta)^2}\right)+\frac{k^{\mu}\eta^{\nu}+k^{\nu}\eta^{\mu}}{(k \cdot \eta)}-\frac{k^{\mu}k^{\nu}\eta^2+\eta^{\mu}\eta^{\nu}k^2}{(k \cdot \eta)^2}.
\end{eqnarray}
if the photon is on the mass shell, $k^2=0$, and the Ward identity is used, $k^{\mu}$ and $k^{\nu}$ terms vanish in the amplitude of monopoles on the mass shell, and hence $\sum_{\gamma's~pol.}\epsilon_{M}(k)^{*\mu}\epsilon_M (k)^{\nu}=-g^{\mu\nu}$.
Therefore, the expression of $\sigma(M+\overline{M} \to 2 ~\mbox{or}~ 3\gamma's)$ is obtained from $\sigma(e^{+}+e^{-} \to 2 \gamma ~\mbox{or}~ 3\gamma)$, by replacing the mass $m_e$ by $m$ and the coupling constant $e$ by $g$. So, $(g\beta)$ depending on the velocity $\beta$ does not appear in our formulation.  On this point, refer to Ref.~\cite{Epele1, Epele2}.

Now we can obtain the decay rates of the para- and ortho-monopoliums as 
\begin{eqnarray}
\Gamma(X_0 \to 2\gamma)&=&4\pi  \left(\frac{g^2}{4\pi}\right)^2 \frac{1}{m^2}\vert \psi_{^1S_0}(R)\vert^2 , \\
\Gamma(X_1 \to 3\gamma)&=&\frac{16(\pi^2-9)}{9}\left(\frac{g^2}{4\pi}\right)^3\frac{1}{m^2}\vert \psi_{^3S_1}(R)\vert^2, 
\end{eqnarray}
where $m$ is the monopole mass and is not the monopolium mass $m_X$. These are the formulae valid at low energy, in the case of $\sqrt{\hat{s}}=m_X  \approx 2m $.  When $\sqrt{\hat{s}}=m_X  \gg 2m $ holds, and the binding energy is large, the following high energy limit formulae should be used:
\begin{eqnarray}
\Gamma(X_0 \to 2\gamma)&=&16\pi \left(\frac{g^2}{4\pi}\right)^2 \frac{1}{m_X^2} (\rho-1) \vert \psi_{^1S_0}(R)\vert^2, \\
\Gamma(X_1 \to 3\gamma)&=&\frac{16}{3} \left(\frac{g^2}{4\pi}\right)^3 \frac{1}{m_X^2} \left[\frac{1}{6} \rho^3-\frac{1}{4}\rho^2+\rho\left(\frac{1}{3}\pi^2-1\right) +3\zeta(3)-\frac{1}{3}\pi^2+3\right] \nonumber \\
&\times& \vert \psi_{^3S_1}(R)\vert^2,
\end{eqnarray}
where $\rho=\ln(\hat{s}/m^2)=\ln(m_X^2/m^2)$, and $\zeta(3)=\sum\limits_{n=1}^{\infty} n^{-3}$=1.202 \cite{Eidelman}. 

For the ortho-monopolium $X_1$, it can decay to quark-antiquark, lepton, and weak boson pairs.  The corresponding decay rates are
\begin{eqnarray}
\Gamma(X_1 \to q\bar{q})&=&3 \cdot \frac{32\pi}{9}\left(\frac{eg}{4\pi}\right)^2 \frac{Q_q^2}{m_X^2}~\vert \psi_{^3S_1}(R)\vert^2, \\
\Gamma(X_1 \to \ell^+\ell^-)&=&\frac{32\pi}{9}\left(\frac{eg}{4\pi}\right)^2 \frac{Q_{\ell}^2}{m_X^2}~\vert \psi_{^3S_1}(R)\vert^2, \\
\Gamma(X_1 \to W^+W^-)&=&\frac{2\pi}{9} \left(\frac{eg}{4\pi}\right)^2\frac{m_X^2}{m_W^4}~\vert \psi_{^3S_1}(R)\vert^2,
\end{eqnarray}
where the electric charge of the quark and lepton is $Q_{q, \ell} ~e$. These formulae are derived for massless initial fermions. So, they have to be applied for the large binding energy case, or $m \ll m_X$ and $m_W  \ll m_X$.  
It is noted that a large number $eg/4\pi$ appears here, but it is not serious, since quarks, leptons and weak bosons don't carry magnetic charge.

The serious contribution to the decay rates comes from the emission of multi-photons, since the (electric or usual) photon can be converted to the magnetic photon and it couples to the monopole with a strong coupling $g$.
Therefore, the total decay rates become
\begin{eqnarray}
\Gamma_\mathrm{tot}(X_0)&=&\sum_{n=\mathrm{even}}\Gamma(X_0 \to n \gamma),  \\
\Gamma_\mathrm{tot}(X_1)&=&\sum_{i} \Gamma(X_1 \to q_i\bar{q}_i)+ \sum_{i}\Gamma(X_1 \to \ell^+_i\ell^-_i)+\Gamma(X_1 \to W^+W^-) \nonumber \\
&+&\sum_{n=\mathrm{odd}}\Gamma(X_1 \to n \gamma).
\end{eqnarray}

It is difficult to estimate the multi-photon emission rate.  However, the contribution of multi-photon emission can reduce the branching ratios $\mathrm{Br}(X_0 \to 2\gamma)$ and $\mathrm{Br}(X_1 \to 3\gamma)$ enormously, so we have to take them into account.  One way is to consider the emitted photons, other than two or three hard photons, to be all soft, namely its energy $E^\mathrm{soft}_{\gamma}$ is less than the energy cut $E^\mathrm{cut}_{\gamma}$.  The energy resolution $\Delta E$ of photons at ATLAS is 
\begin{eqnarray}
\Delta E/E &=& 0.01-0.02 ~~(\mbox{for}~ E > 100~ \mbox{GeV}), \\
&=& (0.1-0.2) /\sqrt{E/\mbox{GeV}} ~~(\mbox{for}~ E < 100~ \mbox{GeV}).
\end{eqnarray}

However, to analyze the data of two photon events, low energy photons are discarded by the energy cutoff  $E^\mathrm{cut}_{\gamma}$, 
\begin{eqnarray}
E^\mathrm{cut}_{\gamma}=25-35~\mbox{GeV},
\end{eqnarray}
which is much larger than the energy resolution \cite{Energy cut1, Energy cut2}.  We choose the energy cut as 35 GeV.  The strategy is, by using the infrared technique, to sum up the emitted but undetectable photons with energy 
\begin{eqnarray}
0 < E^\mathrm{soft}_{\gamma} <  E^\mathrm{cut}_{\gamma}= 35~\mbox{GeV} \ll \frac{m_X}{2}.
\end{eqnarray}

There is another cut of eliminating the photons emitted forward or backward, which is the cut on the pseudo-rapidity $\eta=-\ln \tan(\theta/2)$ ($\theta$ is the scattering angle), namely,
\begin{eqnarray}
\vert \eta \vert ~<~\eta^\mathrm{cut}=2.37 .
\end{eqnarray}
It is rather technical, but we can convert the energy cut and the rapidity cut into nonvanishing photon masses $\lambda$, since the introduction of the photon mass prevents the infrared singularity which occurs at the zero photon energy, as well as the collinear (mass) singularity that occurs when the photon is emitted parallel to the emitting fermion (monopole in this case)
\footnote{The lower bound of the massless photon propagator is identified with the lower bound of massive photon propagator, namely $\left. E^\mathrm{cut}_{\gamma}=\sqrt{\vert \bm{k} \vert^2 + \lambda^2}\right\vert_{\vert \bm{k}\vert=0}$, which gives Eq. (\ref{energy cut}).  Similarly the denominator of the fermion propagator $p\cdot k$ is compared between the case of a massless photon with an angle cut and that of a massive photon without an angle cut. This gives a massless fermion, $1-\cos \theta^\mathrm{cut}=\sqrt{1+(\lambda / \vert \bm{k} \vert)^2}-1$, from which Eq. (\ref{rapidity cut}) is obtained.}.
The conversion is done by the following dictionary:
\begin{eqnarray}
E^\mathrm{cut}_{\gamma} &\Leftrightarrow& \lambda=E^\mathrm{cut}_{\gamma}=35 ~\mbox{GeV}, \label{energy cut} ~\mbox{and}  \\
\eta^\mathrm{cut} &\Leftrightarrow& \lambda=2 E_{\gamma} e^{-\eta^\mathrm{cut}}=0.19 ~E_{\gamma}< 0.19 \left(\frac{m_X}{2}\right). \label{rapidity cut}
\end{eqnarray} 
Upon combining the two cuts, in the case of $m_X \ge $ 750 GeV, the angle cut is stronger, so that the soft photons emitted at LHC satisfy 
\begin{eqnarray}
0< E^\mathrm{soft}_{\gamma} \le 0.19 \left(\frac{m_X}{2}\right). \label{real emission photon}
\end{eqnarray}
The discussion so far has pertained to the emitted photons.

The technique of summing infrared photons is as follows: The contribution of one soft photon emission to the decay width (or the cross section) has an infrared divergence, but the divergence is canceled by the other infrared divergence coming from the radiative corrections of the one soft photon exchange.  As a result the decay width is multiplied by an infrared free factor $\Phi$, called the eikonal factor. This eikonal factor can be easily summed to form an exponential factor, by including multi-photons. Virtual photons should also be soft, since the following approximation is taken in the numerator of the fermion propagator, 
\begin{eqnarray}
\gamma_{\mu} (p+k)^{\mu}+m \approx  \gamma_{\mu} p^{\mu}+m,
\end{eqnarray}
where $p$ is the fermion's (monopole's) momentum and $k$ is the photon's momentum.
This means
\begin{eqnarray}
\vert \bm{k} \vert \ll \{m~ \mbox{and}~ \vert \bm{p} \vert\} \le \frac{1}{2} \sqrt{(2m)^2+ (2\bm{p})^2} \approx \frac{1}{2} \left(2m+ \frac{\bm{p}^2}{m} \right) \approx \frac{1}{2} m_X.
\end{eqnarray}
Therefore, for the virtual soft photon corrections, we have
\begin{eqnarray}
0< E^\mathrm{soft}_{\gamma} \le \frac{1}{2} m_X.  \label{virtual soft photon}
\end{eqnarray}
Then, the cancelation between real emissions and virtual corrections leaves the soft photons in the following energy interval:
\begin{eqnarray}
0.19 \times \frac{1}{2} m_X \le E^\mathrm{soft}_{\gamma} \le \frac{1}{2} m_X.
\end{eqnarray}

If we assume that the monopole has four-momentum $p_1$, the antimonopole has $p_2$ inside the monopolium, and $t=(p_1-p_2)^2=4p_{rel}^2=-4\bm{p}^2_{rel}<0$, then
\begin{eqnarray}
\Phi= -\frac{g^2}{2} \int_{0.19 \times \frac{1}{2} m_X}^{\frac{1}{2} m_X} \frac{d^4k}{(2\pi)^4}\frac{-i}{k^2+i \varepsilon} \left(\frac{p_1^{\mu}}{p_1\cdot k} - \frac{p_2^{\mu}}{p_2\cdot k}\right)^2,
\end{eqnarray}
We can rewrite $\Phi$ using the above dictionary as follows:
\begin{eqnarray}
\Phi &=& \frac{g^2}{2} \left[ \int_{0}^{\infty} \frac{d^4k}{(2\pi)^4}\frac{-i}{k^2-\lambda^2+i \varepsilon} \left(\frac{p_1^{\mu}}{p_1\cdot k} - \frac{p_2^{\mu}}{p_2\cdot k} \right)^2 \right]_{\lambda=0.19 \times \frac{1}{2} m_X}^{\lambda=\frac{1}{2}m_X}\\
&=&\left(\frac{g}{4\pi}\right)^2 \left[-\int_0^1 dy~\frac{y}{y^2+(1-y) (\lambda/m)^2} \right. \nonumber \\ 
& & ~~~~~~~~~~+\left. 4 \left(1-2m^2/t \right)\times \int_0^1 dy~\frac{1}{\sqrt C} \ln \left\vert \frac{\sqrt{C}+y}{\sqrt{C}-y} \right\vert \right] _{\lambda=\lambda=0.19 \times \frac{1}{2} m_X}^{\lambda=\frac{1}{2} m_X}\label{Eikonal} \\
&\approx& \pi^2 \left(\frac{g^2}{4 \pi} \right) \left[\ln^2 \frac{\vert t \vert}{m^2}+4 \ln \frac{\vert t \vert}{m^2} \ln \frac{m}{\lambda} \right]_{\lambda=0.19 \times \frac{1}{2} m_X}^{\lambda=\frac{1}{2} m_X},\label{high energy Eikonal}
\end{eqnarray}
where
\begin{eqnarray}
C=C\left(y, \frac{m^2}{t}, \frac{\lambda^2}{t}\right) =y^2 \left(1-\frac{4m^2}{t}\right) - 4 (1-y) \frac{\lambda^2}{t}.
\end{eqnarray}
The last approximate formula, in Eq. (\ref{high energy Eikonal}), being valid in the high energy limit, is written as a reference, but is useful to estimate $\Phi$ roughly (see for example Ref.~\cite{Landau-Lifshitz-QED}).  We will use Eq. (\ref{Eikonal}), since the high energy limit $\vert t \vert \gg m^2$ does not necessarily work in our case.  This double logarithmic approximation in the high energy limit was used to estimate the total cross section $\sigma(e^{+}e^{-} \to \mbox{multi-photons})$.  Please refer to Ref.~\cite{Eidelman} and references theirein.

Using this eikonal factor, the total decay widths can be written as,
\begin{eqnarray}
\Gamma_\mathrm{tot}(X_0)&=& \Gamma(X_0 \to 2 \gamma) \cosh \Phi, \\
\Gamma_\mathrm{tot}(X_1)&=&\Gamma(X_1 \to 3 \gamma) \cosh \Phi \nonumber \\
&+&\sum_{i} \Gamma(X_1 \to q_i\bar{q}_i)+ \sum_{i}\Gamma(X_1 \to \ell^+_i\ell^-_i)+\Gamma(X_1 \to W^+W^-),
\end{eqnarray}
since we have to add an even number of soft photons to the $2\gamma$ decay or $3\gamma$ decay, the $\cosh \Phi$ factor appears. Now we understand that the branching ratio is roughly given by,
\begin{eqnarray}
\mathrm{Br}(X_0) \approx \mathrm{Br}(X_1) \approx \frac{1}{\cosh \Phi \left(g, t/m^2, m_X/m \right)}. \label{Br}
\end{eqnarray}
Derivation of this suppression factor in the branching ratios is very naive, but is indispensable for the strong coupling dynamics of the monopolium.
\section{A model of a monopole and a monopolium based on the Zwanziger model and lattice gauge theory}
The bound state of the monopoles is formed by the exchange of magnetic photons $\gamma_M$.  However, the coupling of the monopole to the magnetic photon is very strong, so that we adopt a lattice gauge theory approach with a lattice constant $a$ as a UV cutoff \cite{lattice gauge theory1, lattice gauge theory2, lattice gauge theory3}.
The space-time is considered to be a square lattice $n=(n_0, n_1, n_2, n_3)$ ($n_{\mu}$ is an integer) with a lattice constant $a$. The link variables $U^{(A)}_{n\hat{\mu}}$ and $U^{(B)}_{n\hat{\mu}}$ are introduced as usual for the electric and magnetic photons $A_{\mu}(x)$ and $B_{\mu}(x)$, respectively.
\begin{eqnarray}
U^{(A)}_{n\hat{\mu}}=e^{i \{e a A_{\mu}(na)\}}, ~~U^{(B)}_{n\hat{\mu}}=e^{i \{g a B_{\mu}(na)\}},
\end{eqnarray}
and the Wilson loops $W^{(A)}[C]$ and $W^{(B)}[C]$ are defined as the product of the link variables along the loop $C$:

\begin{eqnarray}
W^{(A)}[C]= \prod_{n \in C, ~\mu \parallel C} U^{(A)}_{n\hat{\mu}}, ~~W^{(B)}[C]= \prod_{n \in C, ~\mu \parallel C} U^{(B)}_{n\hat{\mu}}.
\end{eqnarray}
The minimum Wilson loop is given for the boundary curve $C_{n\mu\nu}\equiv \partial P_{n\mu\nu}$ of the minimum plaquette $P_{n\mu\nu}=\mbox{rectangular}~(n, n+\hat{\mu}, n+\hat{\mu}+\hat{\nu}, n+\hat{\nu}, n)$. Then, the Zwanziger action can be written as the lattice gauge theory action in the Euclidean metric:
\begin{eqnarray}
S^{\rm{ZW}}_\mathrm{lattice}&=&-\sum_{n, \nu} \left( \frac{1}{2e^2} W^{(A)} [C_{n\eta\nu}] +\frac{1}{2g^2} W^{(B)} [C_{n\eta\nu}]+ (h.c.) \right) \nonumber \\
&-&\sum_{n, \nu} \frac{1}{2eg}  \left(W^{(A)} [C_{n\eta\nu}]\tilde{W}^{(B)} [C_{n\eta\nu}] - W^{(B)} [C_{n\eta\nu}]\tilde{W}^{(A)} [C_{n\eta\nu}] \right) \nonumber \\
&+&\frac{a^3}{2} \sum_{i, n\nu} \left( \overline{\psi}_{i, n}\gamma_{\nu} \left(U^{(A)}_{n\nu}+U^{(B)}_{n\nu}\right) \psi_{i, n+\nu}-\overline{\psi}_{i, n}\gamma_{\nu} \left(U^{(A)}_{n\nu}+U^{(B)}_{n\nu}\right)^{\dagger} \psi_{i, n-\nu} \right) \nonumber \\
&-& a^4 \sum_{i, n} m_i ~\overline{\psi}_{i, n} \psi_{i, n} ,
\end{eqnarray}
where the dual Wilson loop reads
\begin{eqnarray}
\tilde{W}^{(A,B)} [C_{n\eta\nu}] =-\frac{i}{2} \epsilon_{\eta\nu\lambda\rho}W^{(A,B)} [C_{n\lambda\rho}] ,
\end{eqnarray}
where $\epsilon_{1234}=1$.

In the action, we assume that the magnetic coupling $g$ is strong and , while the electric coupling $e$ is weak and perturbative.  So, in estimating the expectation value of the large Wilson loop $W^{(B)}[C]$, the strong coupling expansion is used \cite{lattice gauge theory1, lattice gauge theory2, lattice gauge theory3}.  We choose $C$ to be a rectangle of length $T$ in time and length $r$ in the space-like direction $\nu$; then we have 
\begin{eqnarray}
\langle W^{(B)}[C] \rangle &=& \frac{\int dU^{(B)}_{n\nu}~ W^{(B)}[C] ~e^{-S^{\rm{ZW}}_\mathrm{lattice}}}{\int dU^{(B)}_{n\nu} ~e^{-S^{\rm{ZW}}_\mathrm{lattice} } }\\
&=& \delta_{\mu\eta} \exp\left(-\ln(2g^2) \frac{T \cdot r}{a^2} \right) \left(1+ \cdots \right),
\end{eqnarray}
where $T \cdot r$ denotes the minimum area of the rectangle $C$, so that the Wilson's area law is realized only if $\mu$ is in the $\eta$-direction.  This point is very important.  Define the potential between a heavy monopole and its antimonopole separated by a distance $r$ to be $V_{M\overline{M}}(r)$.  Then, the potential has a linear term in $r$, if the monopole and antimonopole are separated in the $\eta$-direction;
\begin{eqnarray}
V_{M\overline{M}}(r)= \delta_{r \parallel \eta} \frac{\ln(2g^2)}{a^2} r + \cdots.
\end{eqnarray}
This clarifies the meaning of the special direction $\eta$ that appears in the Zwanziger formulation.  It gives the direction of the Dirac string starting from the monopole.  Therefore, the monopole and antimonopole are connected by the string, starting from the monopole and ending at the antimonopole, which contributes to the linear potential between them.

In addition to this strong coupling contribution, we will add the usual perturbative weak coupling contribution, and so the potential at this stage is
\begin{eqnarray}
V_{M\overline{M}}(r)= -\frac{g^2}{4\pi r} + \frac{\ln(2g^2)}{a^2} r.
\end{eqnarray}
This matches with high precision QCD calculations in which the potential between a quark and antiquark pair is well approximated by the linear plus Coulomb potential \cite{Bali1, Bali2, Bali3}; this potential is good for point-like quarks.  The monopole, however, may not be point-like, and may have an internal structure.  The $ U(1) $ lattice gauge theory is usually considered not to be well defined, since there exists a first-order phase transition between the confinement phase and the perturbative phase \cite{Creutz-Jacobs-Rebbi}, and it obstructs the continuum limit of $a \to 0$.  One way out from this difficulty is to lift the $ U(1) $ theory to $ SU(2) $ theory, or other asymptotic free theory, when we approach to the short distance region.  As is shown by 't Hooft and Polyakov \cite{monopole1, monopole2, monopole3}, the $ U(1) $ monopole was given as a classical solution of $ SU(2) $ gauge theory, which is broken to $ U(1) $ with a triplet Higgs field $\phi^a(x)~(a=1-3)$. Therefore, if we go inside the monopole, the non-Abelian gauge theory may appear.  If this happens we may take the continuum limit properly.  

The 't Hooft-Polyakov monopole is a classical solution of the $ SU(2) $ gauge theory with a triplet Higgs, based on 
\begin{eqnarray}
{\cal L}_{SU(2)\, \mathrm{monopole}}=-\frac{1}{4} (F^a_{\mu\nu})^2+ (D_{\mu}\phi^a)^2-\lambda(\vert\phi \vert^2-v)^2,
\end{eqnarray}
and the following ansatz:
\begin{eqnarray}
A^a_i(x)=v~\epsilon^{aij}\hat{r}^j \frac{1-K(\xi)}{\xi}, ~\phi^a(x)=v~\hat{r}^a \frac{H(\xi)}{\xi},
\end{eqnarray}
where we define a dimensionless parameter $\xi=evr$.
If we take the limit $\lambda \to 0$ while keeping $v \ne 0$, the solution, called the BPS solution, is given analytically \cite{BPS1, BPS2}.  Then, the equations of motion become the first order:
\begin{eqnarray}
\xi \frac{dK}{d\xi}=-KH, ~~\xi \frac{dH}{d\xi}=H-K^2+1,
\end{eqnarray}
from which the solutions read
\begin{eqnarray}
K(\xi)=\frac{\xi}{\sinh \xi}, ~~H(\xi)=\xi \coth\xi-1.
\end{eqnarray}
We want to know the distribution of the magnetic charge inside the monopole, $\xi <1$ or $r< 1/ev$.  The $ SU(2) $ gauge potential and the Higgs field are properly reduced by factors of $1-K(\xi)$ and $H(\xi)$, but the monopole charge is unfortunately not smeared even inside the monopole.  This can be understood from the $ U(1) $ field strength proposed by 't Hooft.  This gauge invariant expression can be rewritten as follows \cite{Arafune-Freund-Goebel}:
\begin{eqnarray}
F_{\mu\nu}=\partial_{\mu} (\hat{\phi}^a A^a_{\nu})-\partial_{\nu} (\hat{\phi}^a A^a_{\mu})-\frac{1}{e}\epsilon^{abc}\hat{\phi}^a \partial_{\mu} \hat{\phi}^b \partial_{\nu} \hat{\phi}^c,
\end{eqnarray}
where $\hat{\phi}^a=\phi^a/\vert \phi \vert$.  From this expression the magnetic charge is found to be a topological number:
\begin{eqnarray}
\Phi_m=\int \bm{B}d\bm{S}=\frac{4\pi}{e} \int d^3x \frac{\partial(\phi^1, \phi^2, \phi^3)}{\partial(x^1, x^2, x^3)}=\frac{4\pi}{e} N,
\end{eqnarray}
where $\Phi_m$ is the magnetic flux, and $N$ is the winding (wrapping) number of the sphere of the Higgs fields ($\bm{\phi}^2=v^2)$ by the  sphere in space ($\bm{x}^2=1$).

We can treat the charged sphere as follows: If the sphere of radius $R$ is uniformly charged, it can be viewed as a point charge $Q(\infty)$ from a long distance, but inside the sphere $(r<R)$ the charge is reduced to $Q(r)=Q(\infty)(r/R)^3$. If we can apply this kind of treatment for the magnetic charge, the Coulomb potential can be relaxed at short distances inside the monopole.  It is not clear, however, whether the magnetic charge can be distributed uniformly inside the sphere-like monopole.

Here, we will consider another possibility in which the magnetic charge is distributed uniformly on the surface of the sphere,  fortunately such a solution exists.  Let's take the following modified gauge and Higgs fields in which the contribution from $r<R$ is cutoff:
\begin{eqnarray}
\tilde{A}^a_i(x)=\theta(r-R) A^a_i(x), ~\tilde{\phi}^a(x)=\theta(r-R)\phi^a(x),
\end{eqnarray}
where $\theta(r-R)$ is a step function $0$ for $r<R$, and $1$ for $r>R$.  Accordingly, $K$ and $H$ are modified:
\begin{eqnarray}
1-\tilde{K}(r)=\theta(r-R) (1-K(r)), ~~\tilde{H}(r)=\theta(r-R)H(r).
\end{eqnarray}
This modification inside the sphere ($r<R$) is allowed, since $K(r)=1$ and $H(r)=0$ is the solution.

The distribution of the $ U(1) $ magnetic charge is easily understood. Since $\phi^a=0$ inside $r<R$, the magnetic flux on the surface of the radius $r$ sphere reads
\begin{eqnarray}
\Phi_m(r)= \frac{4\pi}{e} \theta(r-R),
\end{eqnarray}
which shows that the magnetic charge $4\pi/e$ is distributed uniformly on the surface of the monopole sphere with radius $R$.  The solution satisfies the following equations:
\begin{eqnarray}
\xi \frac{d\tilde{K}}{d\xi}&=&-\tilde{K}\tilde{H} + evR\left(1-K(evR)\right)\delta(\xi-evR), \\
\xi \frac{d\tilde{H}}{d\xi}&=&\tilde{H}-\tilde{K}^2+1+evRH(evR)\delta(\xi-evR),
\end{eqnarray}
which are modified only on the surface, where the magnetic charge is distributed.

Therefore, the magnetic force between the monopole and antimonopole vanishes when one goes inside either respective sphere $(r<R)$.  So, one possibility is to use the following monopole-antimonopole potential $V^{(a)}_{M\overline{M}}(r)$:
\begin{eqnarray}
V^{(a)}_{M\overline{M}}(r)&=& - \frac{g^2}{4\pi r} + \frac{\ln(2g^2)}{a^2} r ~~(\mbox{for}~ r>R), \\
&=&\mbox{const} = - \frac{g^2}{4\pi R} + \frac{\ln(2g^2)}{a^2} R~~ (\mbox{for}~ r<R),
\end{eqnarray}
which is defined by three parameters $g$, $a$ and $R$.  The idea of relaxing the Coulomb potential near the monopole is used by Epele {\it et al.}~\cite{Epele1, Epele2}. They follow Schiff and Goebel and relax the potential through the existence of Dirac strings \cite{Schiff-Goebel1, Schiff-Goebel2}.  Their choice of the potential is 
\begin{eqnarray}
V^{SG}_ {M\overline{M}}(r)&=& - \left(1-e^{-r/R}\right)\frac{g^2}{4\pi r},
\end{eqnarray}
without the linear term.  

Another possibility is to consider the linear potential to be the dominant contribution and the Coulomb potential just playing the role of lowering the potential.  Then, the corresponding potential becomes
\begin{eqnarray}
V^{(b)}_ {M\overline{M}}(r) &=& - \frac{g^2}{4\pi R} + \frac{\ln(2g^2)}{a^2} r ~~(\mbox{for}~ r>R), \\
&=&\mbox{const} = - \frac{g^2}{4\pi R} + \frac{\ln(2g^2)}{a^2} R~~ (\mbox{for}~ r<R).
\end{eqnarray}
The difference between the two potentials is
\begin{eqnarray}
V^{(a)}_ {M\overline{M}}(r)-V^{(b)}_ {M\overline{M}}(r) &=& +\frac{g^2}{4\pi R} -\frac{g^2}{4\pi r}~~(\mbox{for}~ r>R), \\
&=&0~~ (\mbox{for}~ r<R).
\end{eqnarray}

We can also include the spin-dependent hyperfine interaction.  Then, the Hamiltonian $\hat{H}$, which we will use to study the energy and wave function of the lowest s-wave states ($^1S_0$ and $^3S_1$) of the monopolium, is
\begin{eqnarray}
\hat{H}=2m+\hat{H}_\mathrm{NR}+\hat{H}_\mathrm{HF},
\end{eqnarray}
where the nonrelativistic Hamiltonian $\hat{H}_\mathrm{NR}$ and the hyperfine interaction $\hat{H}_\mathrm{HF}$ are defined by
\begin{eqnarray}
\hat{H}_{NR}&=& - \frac{1}{m} \left(\frac{1}{r^2} \frac{d}{dr} r^2 \frac{d}{dr}\right) + V_{M\overline M}(r) , \\
\hat{H}_{HF}&=&\frac{2\pi}{3 m^2}~(\bm{\sigma}_1\cdot \bm{\sigma}_2) ~\bm{\nabla}^2 V_{M\overline M}(r),
\end{eqnarray}
In the case of the Coulomb potential, $g^2 \delta^{(3)}(\bm{r})$ appears from $\bm{\nabla}^2 V(r)$ and the wave function at the origin contributes to the hyperfine interaction.  If we use the potential  $V^{(a)}_ {M\overline{M}}(r)$ or $V^{(b)}_ {M\overline{M}}(r)$, the hyperfine interaction becomes
\begin{eqnarray}
H_{HF}^{(a)}&=&\frac{2\pi}{3m^2}(\bm{\sigma}_1\cdot \bm{\sigma}_2) \left[ \left(\frac{g^2}{4\pi R^2}+\kappa \right) \delta(r-R) + \frac{2\kappa }{r} \theta (r-R) \right], \\
H_{HF}^{(b)}&=&\frac{2\pi}{3 m^2}(\bm{\sigma}_1\cdot \bm{\sigma}_2) \left[ \kappa \delta(r-R) + \frac{2\kappa }{r} \theta (r-R) \right],
\end{eqnarray}
where the string tension is denoted by $\kappa= \ln(2g^2)/a^2$, and $(\bm{\sigma}_1\cdot \bm{\sigma}_2)=-1$ for $X_0$ and $+3$ for $X_1$.  We don't include the relativistic corrections, such as the other Breit terms and the annihilation effect for the hyperfine interaction, since we intend to solve the problem semiclassically.

The mass $m_X$ of the monopole-antimonopole bound state is obtained as the eigenvalue of the total Hamiltonian $\hat{H}$.

\section{Mass and wave function of monopolium} 
In this section we use the potential $V^{(b)}_ {M\overline{M}}(r)$, and obtain the mass and wave function of the monopolium semiclassically.  Schr\"{o}dinger's equation with the linear potential can be solved in terms of the Airy function, but the solution is essentially equal to the semiclassical treatment \cite{Landau-Lifshitz}, so we use the semiclassical method.

For a given nonrelativistic energy $\tilde{E}$, that is an eigenvalue $E_\mathrm{NR}$ of the nonrelativistic Hamiltonian $\hat{H}_\mathrm{NR}$, classical momentum $p_r$ and $\vert p_r \vert$ are given by
\begin{eqnarray}
p_r=\sqrt{m(\tilde{E}-V(r))}, ~~\mbox{and}~~\vert p_r \vert=\sqrt{m(V(r)-\tilde{E})}.
\end{eqnarray}
The turning point $r=r_{*}$ is the point where $p_{r_{*}}=0$.

If we define the wave function by $\psi(r)= \chi(r)/r$, then $\chi(r)$ satisfies the one-dimensional Schr\"{o}dinger equation, and the corresponding semiclassical (WKB) wave function reads
\begin{eqnarray}
\chi(r)_I&=& \frac{C}{2\sqrt{\vert p_r \vert}} e^{-\int^r_{r_{*}} \vert p_r \vert dr}~~\mbox{for}~~r>r_{*}, \\
\chi(r)_{II}&=& \frac{C}{\sqrt{p_r} } \sin \left(-\int_r^{r_{*}} p_r dr + \frac{\pi}{4} \right)~~\mbox{for}~~0<r<r_{*}, 
\end{eqnarray}
and the Bohr-Sommerfeld quantization condition is given by
\begin{eqnarray}
2 \int_0^{r_{*}} p_r ~dr= 2\pi \left(n-\frac{1}{2}\right), \label{BS}
\end{eqnarray}
where $n~(=1, 2, \cdots)$ denotes the radial excitation number, or the principal quantum number.

If we apply this for the potential $V^{(b)}(r)$, then the wave function is analytically given by, 
\begin{eqnarray}
\int^r_{r_{*}} \vert p_r \vert dr &=&\frac{2\sqrt{m}}{3\kappa} \left( \kappa r-(\tilde{E}+g^2/4\pi R) \right)^{3/2}-(r=r_{*})~~\mbox{for}~~r>r_{*}, \\
\int_r^{r_{*}} p_r dr&=& \frac{2\sqrt{m}}{3\kappa} \left((\tilde{E}+g^2/4\pi R)-\kappa r \right)^{3/2}-(r=r_{*})~~\mbox{for}~~R<r<r_{*}, \\
\int_r^{r_{*}} p_r dr&=& \frac{2\sqrt{m}}{3\kappa} \left((\tilde{E}+g^2/4\pi R)-\kappa r \right)^{3/2}(r=R)-(r=r_{*}) \nonumber \\
&+& \sqrt{m} \left((\tilde{E}+g^2/4 \pi R)-\kappa R\right)^{1/2} (R-r)~~\mbox{for}~~0<r<R, 
\end{eqnarray} 
where we use the notation of the string tension as $\kappa=\ln(2g^2)/a^2$.

The Bohr-Sommerfeld quantization condition becomes a third order algebraic equation;
\begin{eqnarray}
\frac{2}{3m\sigma} (p_R)^3 +R (p_R)=\pi (n-1/2),
\end{eqnarray}
for 
\begin{eqnarray}
p_R=\sqrt{m(\tilde{E}+ \frac{g^2}{4\pi R}-\kappa R)}.
\end{eqnarray}
This is the case of one real root and two mutually complex conjugate roots, and so the real root is given by the Cardano formula as
\begin{eqnarray}
\tilde{E}_n=-\frac{g^2}{4\pi R}+\kappa R + \frac{1}{m} \left\{\sqrt[3]{-\frac{q}{2}+\sqrt{\left(\frac{q}{2}\right)^2+\left(\frac{p}{3}\right)^3}} + \sqrt[3]{-\frac{q}{2} - \sqrt{\left(\frac{q}{2}\right)^2+\left(\frac{p}{3}\right)^3}} \right\}^2,
\end{eqnarray}
where 
\begin{eqnarray}
p=\frac{3m}{2} \kappa R,~~\mbox{and}~~q=-\frac{3m}{2} \kappa \pi(n-1/2).
\end{eqnarray}

Here we consider two cases: (Case 1) is that the monopole mass is small compared to the scale of the string tension, and also $R$ is about the same order as the string scale;
\begin{eqnarray}
m \ll \sqrt{\kappa} = \frac{\sqrt{\ln (2g^2)}}{a}, ~~\mbox{and}~~
R = O( 1/\sqrt{\kappa})=O(a),
\end{eqnarray}
Under this condition, $(q/2)^2 \gg (p/3)^3$ , or {\bf [$m\kappa R^3 \ll 9 \pi^2(2n-1)^2/8$]} holds, and we obtain
\begin{eqnarray}
\tilde{E}^{(1)}_n&=&-\frac{g^2}{4\pi R}+\kappa R+\frac{1}{m} (\sqrt[3]{q})^2 \\
&=&-\frac{g^2}{4\pi R}+\kappa R+ \left(\frac{3\pi}{4} \kappa \right)^{2/3} m^{-1/3} (2n-1)^{2/3}.
\end{eqnarray}
The dependence of the energy on the mass $m$ is $(m)^1$ for the Coulomb potential, but is $m^{-1/3}\sigma^{2/3}$ in the linear potential case.\footnote{This dependence $m^{-1/3}$ is utilized in the study of pentaquark mass.  See Eq. (104) in Ref.~\cite{Bando}.}

The other case (Case 2) is that the monopole mass is heavier than the string scale, that is, the case where $(q/2)^2 \ll (p/3)^3$, or {\bf [$m\kappa R^3 \gg 9 \pi^2(2n-1)^2/8$]} holds.  In this case, the energy levels depend on $m^{-1}R^{-2}$:
\begin{eqnarray}
\tilde{E}^{(2)}_n= -\frac{g^2}{4\pi R} +\kappa R +\frac{1}{m} \left(\frac{\pi}{2R} \right)^2 (2n-1)^2.
\end{eqnarray}

The normalization constant $C$ is fixed numerically so as to satisfy the wave function's normalization,
\begin{eqnarray}
4 \pi \int_0^{\infty}  \vert \chi (r) \vert^2 dr = 1,
\end{eqnarray}
and the value of the function at $r=R$ is determined with $C$ or $C_n$ for the $n$th state as 
\begin{eqnarray}
\vert \psi_n(r=R) \vert = \frac{C_n}{R \sqrt{p_R}} \left\vert \sin (p_R \cdot R + \pi/4)\right\vert, 
\end{eqnarray}
where we consider that the monopole and antimonopole collide when the distance $r=R$ in our model, so that the wave function at $r=R$ is used instead of the usual value at $r=0$.

\section{Comparison with experiments, LHC and others} 
In this section, on the basis of the formulae obtained so far, we will compare the numerical estimation of the mass and the decay width of monopoliums ($^1S_0$ and $^3S_1$) , and the production rate with experiments such as LHC and others.   

There are typically two kinds of Solutions, 1 and 2: 

Solution 1 has a large monopole mass about half of the monopolium mass, namely $2m \approx m_X$.  In this case the nonrelativistic energy $\tilde{E}$ is one order of magnitude smaller than the monopole mass, and the hyperfine interaction is also negligibly small. This corresponds to (Case 2) in the previous section. See Table~\ref{tab:monopolium10} in which a few examples of Solution 1 are listed, where $J=0$ and $J=1$ monopoliums are degenerate with a mass of a several TeV. Estimation of the total width $\Gamma_\mathrm{tot}(X) \approx \mathrm{\Gamma}(X \to 2\gamma ~\mbox{or} ~3\gamma) \cosh \Phi$ is performed using formulae (\ref{Eikonal}) and (\ref{Br}) with $\Gamma(X_0 \to 2\gamma)$ (28) and $\Gamma(X_1 \to 3\gamma)$ (29).  However, the very large total widths and the very small lifetimes suggest that our approximation is too naive, and a more reasonable estimation should be developed.  We can say here that the estimation of the total widths and branching ratios are important for the strong coupling monopole dynamics.
\begin{table}[t]
\caption{The derived properties of the monopolium spectrum for masses of $ \mathcal{O}(1-10) $ TeV. 
We use $g=\sqrt{4\pi\cdot137}$, and $R=a$ is taken, while $ a $ and $ m $ are varied.}
\label{tab:monopolium10}
\centering
	\begin{tabular}{| c || c | c || c | c || c | c |} \hline
Spin & 0 & 1 & 0 & 1 & 0 & 1\\ \hline \hline
Mass & \multicolumn{2}{|c||}{$5~\mathrm{TeV}$} & \multicolumn{2}{|c||}{$6~\mathrm{TeV}$} & \multicolumn{2}{|c|}{$7~\mathrm{TeV}$}\\ \hline
$m$~[GeV] & \multicolumn{2}{|c||}{$2624$} & \multicolumn{2}{|c||}{$3149$}  & \multicolumn{2}{|c|}{$3674$} \\ \hline
$\tilde{E}=E_\mathrm{NR}$~[GeV] & \multicolumn{2}{|c||}{$-248$} & \multicolumn{2}{|c||}{$-298$} & \multicolumn{2}{|c|}{$-347$} \\ \hline
$E_\mathrm{HF}$~[GeV] & $-6\cdot10^{-6}$ & $2\cdot10^{-5}$ & $-7\cdot10^{-6}$ & $2\cdot10^{-5}$ & $-8\cdot10^{-6}$ & $2\cdot10^{-5}$\\ \hline
$\Gamma_\mathrm{tot}$~[GeV] & $2 \cdot 10^{15}$ &  $4 \cdot 10^{16}$ & $3 \cdot 10^{15}$ &  $5 \cdot 10^{16}$
& $3 \cdot 10^{15}$ &  $6 \cdot 10^{16}$ \\ \hline
$\tau$~[s] & $3\cdot10^{-40}$ &  $2\cdot10^{-41}$ & $2\cdot10^{-40}$ &  $1\cdot10^{-41}$
& $2 \cdot 10^{-40}$ & $1 \cdot 10^{-41}$ \\ \hline
$a$ [$\mathrm{GeV}^{-1}$] & \multicolumn{2}{|c||}{$1/0.52$} & \multicolumn{2}{|c||}{$1/0.43$} & \multicolumn{2}{|c|}{$1/0.37$}\\ \hline
\end{tabular}
\end{table}
At this stage, the states considered are difficult to observe as a resonance peak at the LHC due to their high masses, but multi-photon emission decays may be observable at future 100 TeV colliders.  Other than collider experiments, they could possibly be observed in astrophysical observations, through an explosion or burst of $\gamma$ rays from the monopoliums.  The number distribution function $f_{\gamma}(n)$ for photons can be very roughly estimated using the eikonal factor as
\begin{eqnarray}
f_{\gamma}(n)=\frac{\Phi^n/n!}{\cosh \Phi} ~~~(n=\mbox{even~for}~ X_0, ~\mbox{odd for~} X_1).
\end{eqnarray}

However, in order to give a definite statement for the number distribution function of photons, a more reasonable estimation of $\Phi$ is necessary.

The second case, Solution 2, is obtained when considering very light monopole masses, for example 870 MeV.  It is found that the large  energy $\tilde{E}$ can lead to a bound state mass of 750 GeV and 1410 GeV for the $ X_0 $ and $ X_1 $ states respectively.  See Table~\ref{tab:monopolium}.

\begin{table}[t]
\caption{Properties of the monopolium and it's decay at a 13 TeV collider.
The parameter choices considered are $g=\sqrt{4\pi\cdot137}$, $a=64/\mathrm{MeV}$, $R=570/\mathrm{MeV}$ 
and monopole mass $m = 870$~MeV, for consistency with the relation in Case 2.}
\label{tab:monopolium}
\centering
\begin{tabular}{| c || c | c |} \hline
Spin & 0 & 1 \\ \hline \hline
Mass & $750~\mathrm{GeV}$ & $1410~\mathrm{GeV}$ \\ \hline \hline
$\sigma$ ($\sqrt{s} = 13~\mathrm{TeV}$) & $0.68$~fb (to 2 $\gamma$) & $0.0025$~ab (to 3 $\gamma$) \\ \hline
Fraction ($\Gamma_i/\Gamma_\mathrm{tot}$) & $0.78$ (2 $\gamma$ decay) & $1$ (3 $\gamma$ decay) \\ \hline
$\tilde{E}=E_\mathrm{NR}$ & $914$~GeV & $914$~GeV \\ \hline
$E_\mathrm{HF}$ & $-164$~GeV & $496$~GeV \\ \hline
$\Gamma_\mathrm{tot}$ & $5$~MeV & $6$~MeV  \\ \hline
\end{tabular}
\end{table}
In the estimation of the production cross section for $X_0$ we have used the formula of Csaki {\it et al.} \cite{Csaki}, 
\begin{eqnarray}
\sigma_{13~\mbox{\scriptsize{TeV}}}=10.8~\mbox{pb} \left(\frac{\Gamma_\mathrm{tot}}{45~ \mbox{\small{GeV}}}\right) \mathrm{Br}^2(X_0 \to \gamma\gamma).
\end{eqnarray}
To attempt to replicate the 750 GeV diphoton excess we make an appropriate choice of parameters to obtain an $ X_0 $ mass of $m_{X_0}=750$ GeV, and an $ X_1 $ mass of $m_{X_1}=1410$ GeV. The production cross section in the 2$\gamma$ channel at 13 TeV LHC is, however, at most 0.68 fb and is one order of magnitude smaller than the recent announcements.  In this study the high energy limit of the decay rates $\Gamma(X_0 \to 2\gamma)$ (30), $\Gamma(X_1 \to 3\gamma)$ (31) are used, and the total widths are found to be 5 MeV, 6 MeV respectively by using Eq. (56).  It is also found that the hyperfine interaction is large because of the small monopole mass, such that the $X_0$ and $X_1$ state masses are strongly separated.   

 For the estimation of the cross section for $X_1$
 we have used the decay width $\Gamma(X_1 \to q\bar{q})$, 
a branching ratio of $X_1$ to $3 \gamma$, $\mathrm{Br} (X_1 \to 3 \gamma) \sim 1$, and the parton luminosity which is derived from graphs by W.J.~Stirling~(MSTW2008NLO PDF)\cite{sterling},
 \begin{eqnarray}
 \left[\frac{1}{s} \frac{d {\cal L}_{q\overline{q}}(\tau)}{d\tau}\right]_{\tau=\frac{m^2_X}{s}} \sim 67~\mathrm{pb}
 \end{eqnarray} 
As for the production cross section of $X_1$, the usual fusion mechanism of quark and antiquark inside a proton is used. 
The value of the cross section 0.0025 ab is rather small.
The parton luminosity $\left[\frac{1}{s} \frac{d {\cal L}_{q\overline{q}}(\tau)}{d\tau}\right]_{\tau=\frac{m_X^2}{s}} $, however,
increases when $s$ becomes larger, so that the production cross section can be several times larger for 33 TeV.
 The integrated luminosity of the order of hundred $(\mathrm{ab})^{-1}$ is necessary at 33 TeV LHC or the future 100 TeV colliders in order to detect 1.4 TeV $X_1$.
\section{Conclusion}
Being inspired by the announcement of the 750 GeV excess in the two photon channel at LHC,  we have considered seriously a manageable model of monopolium, without missing the essence of it.  As a result, we have a number of small but interesting findings which are summarized as follows:

(1) Zwanziger's electric and magnetic dual formulation works properly.  Especially, the meaning of a special direction ($n^{\mu}$ in Zwanziger and $\eta^{\mu}$ in this paper) becomes a little more manifest, since in the lattice formulation, the linear potential appears only in this direction.  This supports that $\eta^{\mu}$ is in the direction of  the string going from the monopole and terminating at the antimonopole.\footnote{We should comment on the special direction $\eta ^{\mu}$ a little more: At the classical level, the classical solution gives a singularity in the direction $\eta ^{\mu}$ along the Dirac string\cite{Zwanziger2}.  At the perturbative level with Feynman rules Eq. (10)-(12), $\eta ^{\mu}$ gives the gauge condition, and hence the amplitudes on the mass shell do not depend on it.  At the non-perturbative level, the potential does depend on $\eta ^{\mu}$ and the energy is stored along its direction connecting monopole and antimonopole, that is, the tensionless Dirac string at the classical level becomes a tension-full string.  So, in order to match the perturbative calculation with the classical and non-perturbative treatments, we have to fix $\eta ^{\mu}$ after the classical solution is obtained. From this observation, a natural way of treating Zwanziger formalism is to relax the frozen property of $\eta ^{\mu}$, by replacing with an external string variable $X^{\mu}(t, \sigma)$, namely, $\eta ^{\mu} \rightarrow \partial X^{\mu}(t, \sigma)/\partial \sigma$.  Here $t$ is a time and $\sigma$ is a parameter representing the extension of the string. Then, the non-perturbative effects may lift the external field of $X^{\mu}(t, \sigma)$ to a dynamical variable with a tension-ful string action.}

(2) For the strong coupling, the lattice version of Zwanziger gives a linear potential.

(3) As a finite sized monopole, the BPS equation for the 't Hooft-Polyakov monopole has another solution in which the $ U(1) $ magnetic charge is distributed uniformly on the surface of a sphere. So, the potential becomes flat inside the monopole.

(4) Now, the monopole-antimonopole potential $V_ {M\overline{M}}(r)$ is linear plus Coulomb, but is cutoff inside, that is,
\begin{eqnarray}
V^{(b)}_ {M\overline{M}}(r) &=& - \frac{g^2}{4\pi R} + \frac{\ln(2g^2)}{a^2} r ~~(\mbox{for}~ r>R), \nonumber \\
&=&\mbox{const} = - \frac{g^2}{4\pi R} + \frac{\ln(2g^2)}{a^2} R~~ (\mbox{for}~ r<R). \nonumber
\end{eqnarray}

The hyperfine interaction is modified depending on the potential which includes the non-perturbative effects.

(5) The mass and wave function of the monopolium ($n ^1S_0$ and $n ^3S_1$) can be given in a tractable manner, by applying the semiclassical method to the potential.  

(6) The branching ratio can be estimated very naively, by using the technique of infrared divergences, leading to $\mathrm{Br}(n ^1S_0 \to 2\gamma)\approx \mathrm{Br}(n ^3S_1 \to 3\gamma)\approx 1/\cosh \Phi$, where $\Phi$ is the calculable eikonal factor that depends on the detection cut for the photon energy and angle.  However, the total width calculated in this way is very big in Table~\ref{tab:monopolium10} (not big in Table~\ref{tab:monopolium}), so that we have to find a more reasonable way to estimate it before a definite statement can be given.

Numerical estimation based on our model of monopolium shows that there are typically two kinds of solutions:

(7) One kind of solution (Solution 1) is the large monopole mass case in which the binding energy and the hyperfine splitting are very small.  So, $X_0$ and $X_1$ are degenerate. A few examples considered give masses of monopoliums to be 5, 6, and 7 TeV.  The signals of the multi-photon explosion or burst may be detected by 100 TeV colliders or by astronomical observations.

(8)  The other kind of solution  (Solution 2) gives an 870 MeV monopole, and the monopole and its antimonopole form a 750 GeV $J=0$ state and a 1.4 TeV $J=1$ state.
The solution can be understood similarly as in QCD, where the so-called small current mass of the monopole is heavily dressed with non-perturbative coats and makes the monopoliums heavy. Unfortunately this solution can't explain the 750 GeV excess at LHC, because the predicted diphoton production rate is one order of magnitude smaller than that of the observed excess. The 1.4 TeV $J=1$ state may provide a target at LHC or future 100 TeV colliders.

This paper has studied how to take into account strong coupling effects in monopole dynamics, especially in the dynamics of monopolium.  We provide a method, in which the bound state is formed non-perturbatively using a strong coupling expansion in lattice gauge theory, while the decay rate is estimated perturbatively, since the final number of emitted photons is fixed.  However, the strong magnetic coupling may give large branching ratios for the multi-photon emission processes, even if they are not intended to be detected.  Overcoming this point is important and is an issue requiring future work.

\section*{Acknowledgments}
The authors give their thanks to Professor Masanori Okawa and Professor Takanori Kono for valuable discussions.


\newpage
\begin{thebibliography}{99}
\bibitem{ATLAS}
ATLAS Collaboration, ATLAS-CONF-2015-081.

\bibitem{CMS}
CMS Collaboration, CMS-PAS-EXO-15-004.

\bibitem{Buttazzo:2015txu}
D.~Buttazzo, A.~Greljo and D.~Marzocca,
Eur.\ Phys.\ J.\ C {\bf 76}, no. 3, 116 (2016).

\bibitem{750 GeV models}
A.~Strumia, Interpreting the 750 GeV digamma excess: a review, arXiv:1605.09401, and references cited therein.

\bibitem{Nambu}
Monopolium (monopole-antimonopole dumb-bell in the standard model):\\
Y.~Nambu, Nucl. Phys. {\bf B130}, 505 (1977).

\bibitem{Yamada et al.}
M.~Yamada, T.~T. Yanagida, and K.~Yonekura. arXiv:1604.07203.

\bibitem{Epele1}
L. N. Epele, H. Fanchiotti, C. A. G. Canal, V. A. Mitsou, and V. Vento, Eur.  Phys. J. Plus {\bf 127}, 60 (2012).
\bibitem{Epele2}
L. N. Epele, H. Fanchiotti, C. A. Canal, and V. Vento, Eur. Phys. J. {\bf C56}, 87 (2008).

\bibitem{Neil1}
  N.~D.~Barrie, A.~Kobakhidze, M.~Talia and L.~Wu,
  Phys.\ Lett.\ B {\bf 755}, 343 (2016).
  
 \bibitem{Neil2}
  N.~D.~Barrie, A.~Kobakhidze, S.~Liang, M.~Talia and L.~Wu,
  arXiv:1604.02803 [hep-ph].

\bibitem{lattice gauge theory1}
K. G. Wilson, Phys. Rev. {\bf D10}, 2445 (1974).

\bibitem{lattice gauge theory2}
J. B. Kogut and L. Susskind, Phys. Rev. {\bf D11}, 395 (1975).

\bibitem{lattice gauge theory3}
A.~Ukawa, ``Kenneth Wilson and lattice QCD", arXiv:1501.04215. 

%
\bibitem{Zwanziger1}
  D.~Zwanziger,
  Phys.\ Rev.\  {\bf 176}, 1489 (1968).
\bibitem{Zwanziger2}
D. Zwanziger, Phys. Rev. {\bf D3}, 880 (1971).

\bibitem{Zwanziger3}
L. V. Laperashvili, H. B. Nielsen, Mod. Phys. Lett. {\bf A14}, 2797 (1999).

\bibitem{Zwanziger4}
L. V. Laperashvili, hep-th/0211227(2002).

\bibitem{Zwanziger5}
C. R. Das, L. V. Laperashvili, H. B. Nielsen, Int. J. Mod. Phys. {\bf A21}, 4479 (2006).

%
\bibitem{Creutz-Jacobs-Rebbi}
M. Creutz, L. Jacobs, and C. Rebbi, Phys. Rev. {\bf D20}, 1915 (1979).

\bibitem{Creutz}
M. Creutz, Phys.Rev. {\bf D21}, 2308 (1980).

\bibitem{Barger-Phillips}
V. D. Barger and R. J. N. Phillips, ``Collider Physics", Westview Press (1996).

\bibitem{Eidelman}
S. I. Eidelman and E. A. Kuraev, Nucl. Phys. {\bf B143}, 353 (1978).

%
\bibitem{Energy cut1}
ATLAS Collaboration, arXiv: 1606.3833.

\bibitem{Energy cut2}
ATLAS Collaboration, Eur. Phys. J. {\bf C74}, 3071 (2014).

\bibitem{Landau-Lifshitz-QED}
L. D. Landau and E. M. Lifshitz, ``Quantum Electrodynamics", Pergamon Press (1979).

\bibitem{Bali1}
G. S. Bali, K. Schilling, Phys. Rev. {\bf D47}, 661 (1993).

\bibitem{Bali2}
G. S. Bali, K. Schilling, A. Wachter, in ``Proc. of 2nd Int. Conf. on Quark Confinement and the Hadron Spectrum", hep-ph/9611226.

\bibitem{Bali3}
G. S. Bali, Phys. Rept. {\bf 343}, 1 (2000).

\bibitem{monopole1}
G. 't Hooft,  Nucl. Phys. {\bf 79}, 275 (1974).

\bibitem{monopole2}
A. M. Polyakov, JETP Lett. {\bf 20}, 194 (1974).

\bibitem{monopole3}
James N. Frey, http://www.phys.ufl.edu/~fry/7097/monopole.pdf.

\bibitem{sugamoto}
A.~Sugamoto, Phys. Lett. {\bf B127}, 75 (1983).

\bibitem{Dirac1}
P. A. M. Dirac, Proc. Roy. Soc. {\bf A133}, 60 (1931).

\bibitem{Dirac2}
P. A. M. Dirac, Phys. Rev. {\bf 74}, 817 (1948).

\bibitem{Dirac3}
J. Schwinger, Phys. Rev. {\bf 144}, 1087 (1966).

\bibitem{Noguchi-Sugamoto}
A.~Noguchi and A.~Sugamoto, TSPU Bulletin {\bf 44N7}, 59 (2004) (hep-th/040845).

\bibitem{BPS1}
E. B. Bogomol'nyi, Yad. Fiz. {\bf 24}, 861 (1976) [Sov. J. Nucl. Phys. {\bf 21}, 449 (1976)].

\bibitem{BPS2}
M. K. Prasad and C. M. Sommerfield, Phys. Rev. Lett. {\bf 35}, 760 (1975).

\bibitem{Arafune-Freund-Goebel}
J.~Arafune, P. G. O. Freund, C. J. Goebel, J. Math. Phys. {\bf 16}, 433 (1975).

\bibitem{Schiff-Goebel1}
L. I. Schiff, Phys. Rev. {\bf 160}, 1257 (1967).

\bibitem{Schiff-Goebel2}
C. J. Goebel, ``Quanta, Essays in Theoretical Physics", eds.  P. G. O. Fruend, C. J. Goebel, and Y. Nambu (Chicago, 1996).

\bibitem{Landau-Lifshitz}
L. D. Landau and E. M. Lifshitz, ``Quantum Mechanics", Pergamon Press (1965).

\bibitem{Bando}
M. Bando, T. Kugo. A. Sugamoto, and S. Terunuma, Prog. Theor. Phys. {\bf 112}, 325 (2004).

\bibitem{sterling}
W. J. Stirling, private communication.

\bibitem{Csaki}
C. Csaki, J. Hubisz, S. Lombardo, and J. Terning, Phys. Rev. {\bf D93}, 095020 (2016).


\end{thebibliography}
%

%
%

\end{document}